\documentclass[tradiabstract]{aa} % for a referee version
\usepackage{graphicx}
\usepackage{txfonts}
\usepackage{natbib}
\begin{document}
   \title{Magnetic activity in the photosphere of CoRoT-Exo-2a\thanks{Based on observations obtained with CoRoT, a space project operated by the French Space Agency, CNES, with partecipation of the Science Programme of ESA, ESTEC/RSSD, Austria, Belgium, Brazil, Germany, and Spain.}} 
   \subtitle{Active longitudes and short-term spot cycle in a young Sun-like star}

%   \subtitle{}

   \author{A.~F.~Lanza
          \inst{1}
          \and
          I.~Pagano\inst{1} \and G.~Leto\inst{1} \and S.~Messina\inst{1} 
        \and S.~Aigrain\inst{2} \and R.~Alonso\inst{3} \and M.~Auvergne\inst{4} \and A.~Baglin\inst{4} 
       \and P.~Barge\inst{3} \and A.~S.~Bonomo\inst{1,3,5} \and P.~Boumier\inst{6} \and A.~Collier Cameron\inst{7} \and M.~Comparato\inst{1,5} 
\and G.~Cutispoto\inst{1} \and J.~R.~De~Medeiros\inst{8} 
\and B.~Foing\inst{9} \and A.~Kaiser\inst{10} \and C.~Moutou\inst{3} \and P.~S.~Parihar\inst{11} 
\and A.~Silva-Valio\inst{12} \and W.~W.~Weiss\inst{10}
          }

   \institute{INAF-Osservatorio Astrofisico di Catania, Via S.~Sofia, 78, 95123 Catania, Italy\\
              \email{nuccio.lanza@oact.inaf.it}
         \and
    School of Physics, University of Exeter, Stocker Road, Exeter, EX4 4QL, United Kingdom 
   \and  
Laboratoire d'Astrophysique de Marseille (UMR 6110),
Technopole de Ch\^{a}teau-Gombert,
38 rue Fr\'ed\'eric Joliot-Curie,
13388 Marseille cedex 13, France
    \and 
    LESIA, CNRS UMR 8109, Observatoire de Paris, 5 place J. Janssen, 92195 Meudon, France     
   \and 
    Dipartimento di Fisica e Astronomia, Universit\`a degli Studi di Catania, Via S. Sofia, 78, 95123 Catania, Italy
    \and 
 Institut d'Astrophysique Spatiale (IAS), Batiment 121, F-91405 Orsay, France; Universit\'e Paris-Sud 11 and CNRS (UMR 8617)   
    \and 
    School of Physics and Astronomy, University of St. Andrews, 
    North Haugh, St Andrews, Fife Scotland KY16 9SS   
    \and 
    Departamento de F\'{\i}sica, Universidade Federal do Rio Grande do Norte, 59072-970 Natal, RN, Brazil
   \and 
   ESA/ESTEC/SRE-S, Postbus 299, 2200 AG Noordwijk, The Netherlands
   \and 
   Institut f\"ur Astronomie, Universit\"at Wien, T\"urkenschanzstra{\ss}e 17, 1180, Vienna, Austria
   \and 
   Indian Institute of Astrophysics, Block II, Koramangala, Bangalore, 560034 India 
   \and 
  CRAAM, Universidade Presbiteriana Mackenzie, Rua da Consola\c{c}\~{a}o, 896, S\~{a}o Paulo, SP, 01302, Brazil   }

   \date{Received ; accepted }

% \abstract{}{}{}{}{} 
% 5 {} token are mandatory
 
  \abstract
  % context heading (optional)
  % {} leave it empty if necessary  
   {The space experiment CoRoT has recently detected transits by a hot Jupiter across the disk of an active G7V star (CoRoT-Exo-2a) that can be considered as a good proxy for the Sun at an age of 
approximately 0.5 Gyr.}
  % aims heading (mandatory)
   {We present a spot modelling of the optical variability of the star during 142 days of uninterrupted observations performed by CoRoT with unprecedented photometric precision.}
  % methods heading (mandatory)
   {We apply spot modelling approaches previously tested in the case of the Sun by modelling total solar irradiance variations, a good proxy for  the optical flux variations of the Sun as a star. The best results in terms of mapping of the surface brightness inhomogeneities are obtained by means of maximum entropy regularized models. To model the light curve of CoRoT-Exo-2a, we take into account both the photometric effects of cool spots as well as  those of solar-like faculae, adopting solar analogy.}
  % results heading (mandatory)
   { Two active longitudes initially on opposite hemispheres are found on the photosphere of CoRoT-Exo-2a with a rotation period of $4.522 \pm 0.024$ days. Their separation changes by $\approx 80^{\circ}$ during the time span of the observations.  From this variation, a relative amplitude of the surface differential rotation lower than $\sim 1$ percent is estimated. Individual spots form within the active longitudes and show an angular velocity about $\sim 1$ percent smaller than that of the longitude pattern. The total spotted area shows a cyclic oscillation with a period of $28.9 \pm 4.3$ days, which is close to 10 times the synodic period of the planet as seen by the rotating active longitudes. We discuss the effects of solar-like faculae on our models,  finding  indication of a facular contribution to the optical flux variations of CoRoT-Exo-2a being significantly smaller than in  the present Sun. }
  % conclusions heading (optional), leave it empty if necessary 
  {The implications of such results for the internal rotation of CoRoT-Exo-2a are discussed  on the basis of solar analogy. A possible  magnetic star-planet  interaction is suggested by the cyclic variation of the spotted area. Alternatively, the $28.9$-d cycle may be related to Rossby-type waves propagating in the subphotospheric layers of the star.}


   \maketitle
%
%________________________________________________________________

\section{Introduction}

The CoRoT satellite is a space experiment devoted to asteroseismology and  extrasolar planet search by the observation of planetary transits across the discs of their parent stars
\citep{Baglinetal06}. One of the first planets detected is CoRoT-Exo-2b, a hot Jupiter orbiting a main-sequence G7 star with a period of 1.743 days. The planet's characteristics and the physical parameters of the star have been determined by combining CoRoT photometry with ground-based spectroscopy \citep[][]{Alonsoetal08,Bouchyetal08}. In the present work we address the magnetic activity of the star as revealed by the rotational modulation of its broad-band optical flux induced by photospheric brightness inhomogeneities. The short rotation period ($\sim 4.5$ days) and the sizeable convective envelope of CoRoT-Exo-2a promote a vigorous dynamo action that amplifies and modulates its magnetic fields. Magnetic flux tubes emerging to the surface give rise to cool spots and bright faculae in the photosphere of this star whose mass, radius and effective temperature are comparable to those of the Sun. \citet{Bouchyetal08} suggest that the star is still close to the ZAMS, possibly younger than 0.5 Gyr, representing a unique laboratory to study the magnetic activity of the young Sun, at a level of flux variations one order of magnitude greater than the present Sun. 

We present the results of a spot modelling of CoRoT-Exo-2a based on the methods already applied to model the bolometric variability of the Sun as a star \citep{Lanzaetal03,Lanzaetal04,Lanzaetal07}.
The uninterrupted time series and the high photometric precision of CoRoT observations allow us
to map the longitudinal distribution of the active regions in the photosphere of the star and follow their evolution continuously along a period of five months which has not been previously possible with ground-based observations. Our approach is based on the modelling of stellar flux variations outside transits. The possibility  to map the spots located on the band of the star occulted by the planet  during its transits,  by means of eclipse mapping techniques
\citep[e.g,][]{Silva03,Pontetal07}, 
 will be the subject of another investigation that will benefit from the present study to put its results in a broader context. 

\section{Observations}

CoRoT-Exo-2a was observed during the first long run from May 16 to October 5, 2007 with a time sampling of 512 s during the first week and 32 s thereinafter. CoRoT is on a low-Earth polar orbit with a period of 1.7 hr \citep{Auvergneetal08}, so the observations during the passage of the satellite across the South Atlantic Anomaly of the Earth magnetosphere are systematically affected by proton hits and must be discarded. These interruptions usually do not exceed $15-20$ percent of each orbit and have no significant impact for our study since the modification of the photospheric pattern of an active star occurs on timescales of the order of one day or longer.  Data were corrected for the CCD zero offset and gain. The contribution of the background light was estimated and corrected by observing a window of 100 pixels close to the target. 
After reduction and correction for the known instrumental effects, those observations provide us with one of the highest quality CoRoT light curves. 

The passband of the photometric data used in the present study ranges from 300 to 1100 nm. We preferred to use CoRoT white light data instead of the three colours because of the significantly higher signal-to-noise ratio, reduced impact of the hot pixels, and long-term stability of the signal. 
Transits were removed by means of the ephemeris and parameters of \citet{Alonsoetal08}. To eliminate residual outliers, we disregarded all data points that deviated from the mean more than 4.2 standard deviations of the whole data set. 
{Such a broad limit was adopted to initially include data points that have a deviation from the mean comparable with the depth of the transits. }
Then we computed a filtered version of the light curve by means of a sliding median boxcar filter with a boxcar extension equal to one orbital period of the satellite, i.e., 6184 s \citep[cf. ][]{Auvergneetal08}. This filtered light curve was subtracted from the original light curve and all the points deviating more than 3 standard deviations of the residuals were discarded.
In this way, 0.77 percent of the data initially validated by the CoRoT N2 reduction pipeline were discarded. 
 Finally, we computed normal points by binning the data on time intervals having the duration of the orbital period of the satellite, getting a light curve consisting of 1945 normal points that cover 142.006 days. Each normal point is obtained by averaging  $\sim 193$ observations with a 32-s sampling and its mean standard error is 
$1.03 \times 10^{-4}$ in relative flux units. 

As pointed out by \citet{Alonsoetal08}, a faint apparent companion located $4^{\arcsec}$ apart  falls completely inside the photometric mask of CoRoT-Exo-2a contributing to  5.6$\pm$0.3 percent of the total flux. 
This fraction of the median flux has been subtracted and the
light curve  normalized to its maximum value of 701846 (e$^{-}$/32 s), observed at
HJD = 2454374.2861, that we assume to
represent the unspotted flux level of the star, whose true value is unknown.

\section{Spot modelling}

To fit the high precision light curves of solar-like stars provided by space-borne photometers, we  apply two approaches that have been tested using observations of the total solar irradiance, i.e., of the Sun as a star, by \citet{Lanzaetal07}. The first is based on the use of three active regions, containing both cool spots and warm faculae, to fit the rotational modulation of the flux, and of a uniform component to fit the variation of the mean light level produced by a uniform background  of active regions. This model will be hereinafter referred to as the 3-spot model and it is described in detail in \citet{Lanzaetal03,Lanzaetal07}. Its main advantage is the limited number of free parameters. When the inclination of the stellar rotation axis is fixed, as in our case (see Sect.~\ref{model_param}), we  fit the model to the observed flux variations by varying the area, longitude and latitude of the three active regions, their rotation period and the mean flux level, having in total 11 free parameters. The  limb darkening parameters  and the spot and facular contrasts are determined by fixing the stellar, spot and facular temperatures \citep[see][]{Lanzaetal04,Lanzaetal07}. The ratio $Q$ of the facular to the spotted area in the active regions is determined by minimizing the $\chi^{2}$ of the best fit to the entire light curve, according to the method suggested by \citet{Gondoin08} (see Sect.~\ref{model_param}). 

A second method applied to fit the light curves of solar-like stars is based on the use of a continuous distribution of active regions to reproduce the complex surface patterns observed on the Sun close to the maximum of the eleven-year cycle or on stars that are more active than the Sun itself \citep[cf., e.g., ][]{Kovarietal04}. The star is subdivided into a large number of surface elements, in our case 200  squares of side $18^{\circ}$, with  each element containing unperturbed photosphere, cool spots and facular areas. The fraction of an element area covered by cool spots is indicated by the filling factor $f$,  the fractional  area of the faculae is $Qf$ and the fractional area of the unperturbed photosphere is $1-(Q+1)f$. We fit the light curve by changing the value of $f$ over the surface of the star, while $Q$ is held uniform and constant. Even fixing the rotation period, the inclination, and the spot and facular contrasts \citep[see ][ for details]{Lanzaetal07}, the model has 200 free parameters and suffers from  non-uniqueness and instability. To find a unique and stable spot map, we apply maximum entropy regularization (hereinafter ME), as described in \citet{Lanzaetal07}, by minimizing a functional 
$\Theta$ which is a linear combination of the $\chi^{2}$ and  the entropy functional $S$, i.e.:
\begin{equation}
\Theta = \chi^{2} ({\vec f}) - \lambda S ({\vec f}),
\end{equation}
where ${\vec f}$ is the vector of the filling factors of the surface elements, $\lambda > 0$  a Lagrangian multiplier determining the trade-off between light curve fitting and regularization, and the expression of $S$ is given in \citet{Lanzaetal98}. To fix the optimal value of the Lagrangian multiplier, we compute the mean of the residual for the regularized best fit $\mu_{\rm reg}$ and compare it with the standard error obtained for the unregularized best fit, i.e., $\epsilon_{0}=\sigma_{0}/\sqrt{N}$,
where $\sigma_{0}$ is the standard deviation of the residuals obtained for $\lambda=0$ and $N$ is the number of normal points in the data set.  
 We fix the optimal value of $\lambda$ by requiring that $|\mu_{\rm reg}|= \epsilon_{0}$, i.e., we iterate on $\lambda$ until the mean of the residuals of the ME best fit is equal in modulus to one standard error of the residuals of the unregularized model. 

Given the  evolution of the sunspot configuration on the Sun, it is possible to obtain a good fit of the irradiance changes,  both with the 3-spot and the ME models, only for a limited time interval $\Delta t_{\rm f}$, not exceeding 14 days that is the lifetime of the largest sunspot groups dominating the irradiance variation. In the case of other active stars, the value of        $\Delta t_{\rm f}$ must be determined from the observations themselves, looking for the maximum data extension that allows for a good fit with the applied models (see Sect.~\ref{model_param} for CoRoT-Exo-2a). 

The activity level of CoRoT-Exo-2a is significantly higher than that of the Sun, as indicated by the amplitude of the flux modulation which is $\sim 20$ times greater than that of the Sun at the maximum of the eleven-year cycle \citep[cf. ][]{Alonsoetal08}. Therefore, we expect that its light variations are dominated by cool spots, as suggested by { \citet{Radicketal98} } and \citet{Lockwoodetal07}. In fact, the light curve variations can be accurately modelled by considering only cool spots, as we shall show in Sect.~\ref{results}, but the possible role of solar-like faculae cannot be excluded a priori, so we shall consider also models with a facular component. In Sect.~\ref{model_param}, we shall use the 3-spot model, that has a  number of free parameters much smaller than that of the ME model, to estimate the best value of the facular-to-spotted area ratio $Q$. 

The inclination of the rotation axis of CoRoT-Exo-2a is almost $90^{\circ}$  \citep{Alonsoetal08}, thus the duration of the transit of a given spot across the stellar disc is independent of its latitude in the hypothesis that surface differential rotation is negligible. In the case of the Sun, the angular velocity difference produced by the latitudinal differential rotation in the sunspot belt (i.e.
$\pm 40^{\circ}$ from the equator) is only $\sim 4$ percent, i.e. too small to be appreciated given the limited lifetimes of the sunspot themselves. Therefore, even if the amplitude of the differential rotation in CoRoT-Exo-2a were  comparable to that of the Sun, we do not expect an  observable dependence of the spot transit time on the latitude. In other words, no information on spot latitude can be extracted from the light curve modelling. On the other hand, the spot longitudes and the variation of the total spotted area during the 142 days of observations can be reliably extracted from the light curve. Our extended tests conducted on the total solar irradiance variation along cycle 23 show that ME models are superior to the 3-spot models to derive the distribution of the spotted area vs. longitude and the variation of the total spotted area. The longitude of the largest spot groups can be determined with an accuracy of at least $15^{\circ}-20^{\circ}$ and the resolution in longitude, i.e., the capability of separating neighbour groups of comparable area, is of $\sim 50^{\circ}-60^{\circ}$ \citep[cf. ][]{Lanzaetal07}. If solar-like faculae  contribute to the light modulation of the star, the centre-to-limb variation of their brightness contrast may systematically affect the derivation of the active region longitudes, as discussed by \citet{Lanzaetal07}. Therefore, another motivation for computing models with a facular contribution is to estimate the systematic errors of active region longitudes. 

\section{Model parameters}
\label{model_param}

%%%%%%%%%%%%%%%%%%%%%%%%%%%%%%%%%%%%%%%%%%%%%%%%%%%%%%%%%%%%%%%%%
\begin{figure}[t]
\includegraphics[width=8cm,height=6cm]{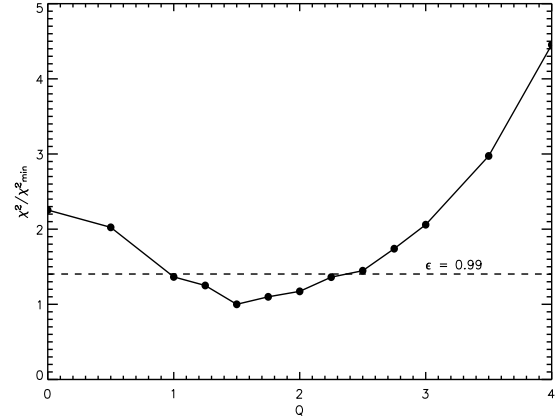} % qfac.ps 
\caption{The ratio of the $\chi^{2}$ of the composite best fit to the entire time series of CoRoT-Exo-2a
to its minimum value vs. the parameter $Q$, i.e., the ratio of the area of the faculae to that of the cool spots in active regions. The horizontal dashed line indicates the 99 percent confidence level for $\chi^{2}/\chi_{\rm min}^{2}$, determining the interval of the acceptable $Q$ values.}
\label{qratio}
\end{figure}
%%%%%%%%%%%%%%%%%%%%%%%%%%%%%%%%%%%%%%%%%%%%%%%%%%%%%%%%%%%%%%%%%
%%%%%%%%%%%%%%%%%%%%%%%%%%%%%%%%%%%%%%%%%%%%%%%%%%%%%%%%%%%%%%%%%
\begin{figure*}[]
\centerline{
\includegraphics[width=14cm,angle=90]{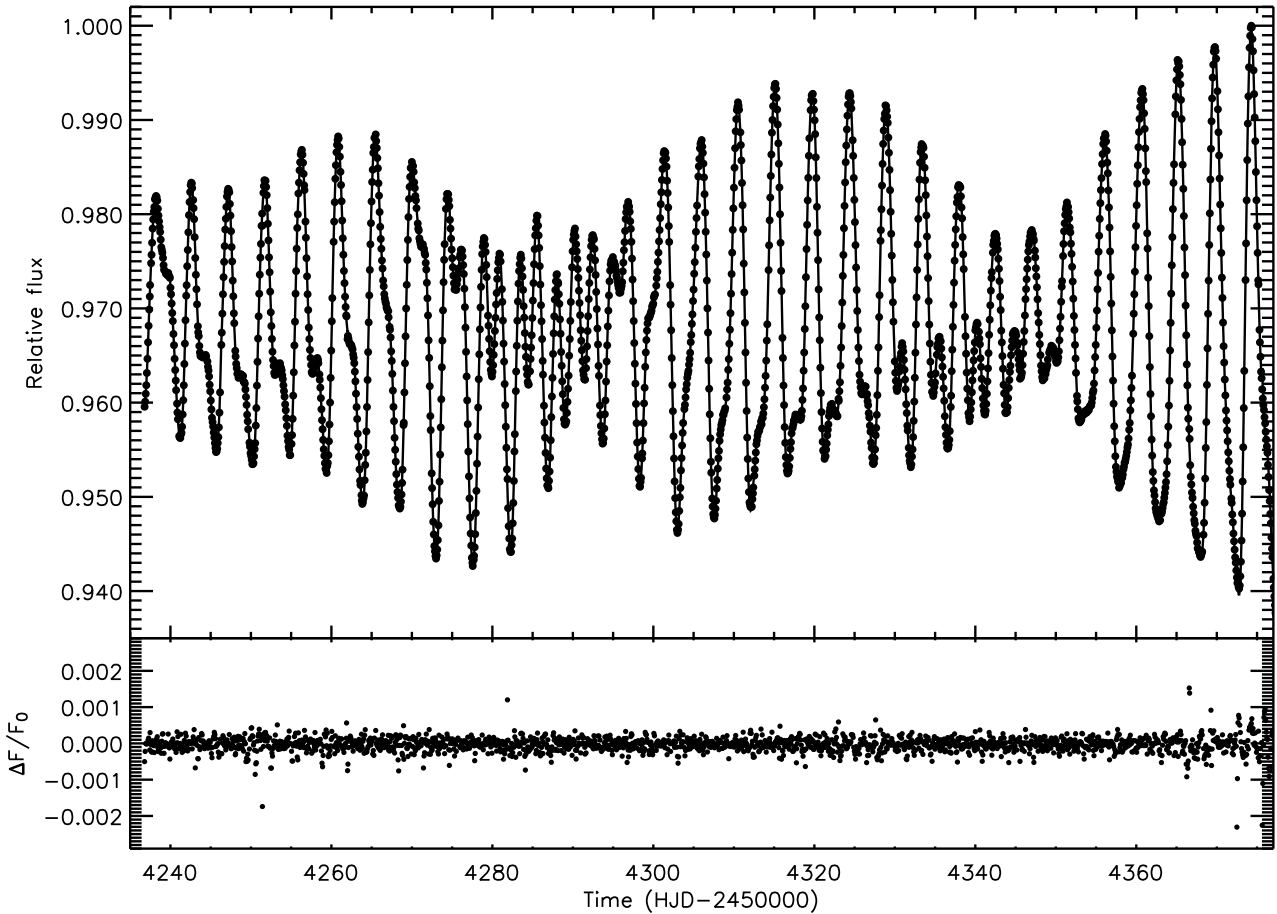} % gen-fit.ps 
}
\caption{Upper panel: Out-of-transit observations of CoRoT-Exo-2a (filled dots) versus time with the best fit obtained with the ME model including only cool spots (solid line); lower panel: the residuals of the best fit in relative flux units versus time. }
\label{best_fit}
\end{figure*}
%%%%%%%%%%%%%%%%%%%%%%%%%%%%%%%%%%%%%%%%%%%%%%%%%%%%%%%%%%%%%%%%%

The basic stellar parameters are taken from \citet{Alonsoetal08} and \citet{Bouchyetal08} and are listed in Table\ref{model_param_table}. We adopt the same quadratic limb-darkening law of \citet{Alonsoetal08} with the parameters derived from the fitting of the planetary transit profile. The inclination of the stellar rotation axis is constrained by the Rossiter-McLaughlin effect \citep[see ][]{Bouchyetal08}, so we can assume that it is perpendicular to the orbital plane of the planet. We assume an isophotal wavelength of the CoRoT passband of 500 nm to take into account the surface distortion and the 
gravity darkening effect induced by the centrifugal force. Considering a rotation period of $4.52$ days (see below) the relative difference between the polar and the equatorial axes $\epsilon$,  evaluated with the Roche model,  is 
$2.48 \times 10^{-4}$ which may induce a  flux variation about one order of magnitude smaller for a spot coverage of 10 percent of the stellar surface.  
The time-dependent tidal distortion due to the close-by planet, having a mass of 3.3 Jupiter masses, has been computed with the model of \citet{Pfahletal08} and induces a negligible flux variation with a relative amplitude of $\sim 3 \times 10^{-5}$. The flux variation associated with the Doppler shift produced by the star's reflex motion has a relative amplitude of only $8.0 \times 10^{-6}$ \citep{LoebGaudi03}. The light reflected by the planet reaches $1.0 \times 10^{-4}$ in relative flux units if we adopt a geometric albedo of 13 percent, that is quite high in view of the hot Jupiter atmospheric models computed by, e.g., \citet{Sudarskyetal00}, \citet{Seageretal00}, \citet{Burrowsetal08}. They are supported by the upper limit for the albedo of HD~209458b recently derived by \citet{Roweetal06,Roweetal07}.     

The rotation period has been found by means of the Lomb-Scargle periodogram \citep{HorneBaliunas86} and is $P_{\rm rot} = 4.52 \pm 0.14$ days, with the uncertainty dominated by the finite time extension of the data set. Since starspots are the tracers used to find the rotation period and considering that they evolve  and migrate in longitude versus time, a refinement of the value given by the Lomb-Scargle periodogram can be found by means of the spot modelling itself. 
After some tests, we found that a value of the rotation 
period of $4.5221$ days minimized the longitudinal migration of the active  longitudes where new starspots were forming during the interval of the observations, thus we adopted this value as the rotation period for our  modelling (cf. Sect.~\ref{results}).   

The maximum time interval $\Delta t_{\rm f}$ that can be fitted both with the 3-spot and the ME models turned out to be $ \sim 3.2$ days. We adopted a value $\Delta t_{\rm f} = 3.15611$ days in order to have 45 equal intervals covering the entire time series. Note that even if each fitted data subset is only $\sim 70$ percent of a rotation period, the results obtained with the ME spot modelling are fully meaningful, as confirmed by the tests performed in the case of the Sun where the models were computed using subsets with an extension of only 
$\sim 50$ percent of a solar rotation \citep[cf. ][ for details]{Lanzaetal07}. 

The brightness contrast of the spots $c_{\rm s}=0.665$ was assumed to be the same of the sunspot bolometric contrast  \citep{Chapmanetal94} and is independent of the limb angle, as discussed by \citet{Lanzaetal03,Lanzaetal04}. It corresponds to a spot temperature $\sim 540$ K below that of the unperturbed photosphere that, in the case of the Sun, accounts for the effects of the extended penumbra in large sunspot groups. { Considering stars with  photometric variations of a few 0.1 magnitudes, \citet{Berdyugina05} found a relationship between the temperature of the spots and that of the unperturbed photosphere. In the case of CoRoT-Exo-2a, it would imply a temperature deficit as large as $\sim 1700$ K, corresponding to  $c_{\rm s} \sim 0.21$. The real temperature deficit of the spots on our star is probably somewhere in between the extreme values considered above, given that its level of activity is intermediate between that of the Sun and those of the most active stars. We prefer to adopt a strict solar analogy for the present analysis assuming that the active regions of CoRoT-Exo-2a consist of spots analogous to sunspots, but with a larger area or filling factor. However, we briefly discuss at the end of Sect.~\ref{results} how our results are affected if the spot contrast is changed or  is not uniform over the photosphere of the star.}

In the case of the models including the facular contribution, we assume the same 
facular contrast  of the Sun $c_{\rm f}=0.115$, with the same dependence on the limb angle, giving a zero contrast at the centre of the disc and a maximum contrast at the limb \citep{Lanzaetal04}.

The best value of the area ratio $Q$ between the faculae and the spots in each active region has been estimated by means of the 3-spot model. In Fig.~\ref{qratio}, we plot the ratio $\chi^{2}/ \chi^{2}_{\rm min}$ of the 
total $\chi^{2}$ of the composite best fit to the entire time series to 
its minimum value $\chi^{2}_{\rm min}$, versus $Q$. The horizontal dashed line indicates the 99 percent confidence level as derived from the F-statistics \citep[e.g., ][]{Lamptonetal76}. The best value of $Q$ turns out to be $Q=1.5$, with an acceptable range extending from $\sim 1$ to $\sim 2.5$. For comparison, the best value in the case of the Sun is $Q_{\odot}=9$, indicating a lower relative contribution of the faculae to the total light variation in late-type stars more active than the Sun, as suggested by \citet{Gondoin08}. Considering the colour light curves of CoRoT-Exo-2a, we searched for the characteristic double light maxima produced by the transits of  faculae across the disc of the star, particularly evident at shorter wavelengths, as it is often  observed in the Sun. We found no evidence of such  facular signatures, confirming that the facular contribution, if present, is indeed significantly smaller than in the case of the Sun. 
{This may suggest that for a low filling factor of the magnetic flux tubes, facular brightness enhancement would dominate, whereas when the filling factor exceeds a certain threshold, convective heat transport is inhibited and dark spots appear with an increasingly relative contribution \citep[cf. also ][]{SolankiUnruh04}.} 

%%%%%%%%%%%%%%%%%%%%%%%%%%%%%%%%%%%%%%%%%%%%%%%%%%%%%%%%%%%%%%%%%%%%%%%%%%%%%%%%
\begin{table}
\caption{Parameters adopted for the spot modelling}
\begin{tabular}{lrr}
\hline
 & & \\
Parameter &  & Ref.$^{a}$\\
 & & \\ 
\hline
 & &  \\
Star Mass ($M_{\odot}$) & 0.97 & A08  \\
Star Radius ($R_{\odot}$) & 0.90 & A08  \\
$T_{\rm eff}$ (K) & 5625 &  B08 \\ 
Limb darkening $u_{\rm a}$ & 0.41 & A08 \\
Limb darkening $u_{\rm b}$ & 0.06 & A08 \\ 
$P_{\rm rot}$ (d) & 4.5221 & L08 \\
$\epsilon$ & $2.48 \times 10^{-4}$ & L08 \\ 
Inclination (deg) & 87.84 & A08  \\
$c_{\rm s}$  & 0.665 & L08 \\
$c_{\rm f}$  & 0.115 & L04 \\ 
$Q$ & 0.0, 1.5  & L08 \\ 
$\Delta t_{\rm f}$ (d) & 3.15611 & L08 \\ 
& &   \\
\hline
\label{model_param_table}
\end{tabular}
~\\
$^{a}$ References: A08: Alonso et al. (2008); B08: Bouchy et al. (2008); L04: Lanza et al. (2004); L08: present study.
\end{table}
%%%%%%%%%%%%%%%%%%%%%%%%%%%%%%%%%%%%%%%%%%%%%%%%%%%%%%%%%%%%%%%%%%%%%%%%%%

\section{Results}
\label{results}

The sequence of best fits obtained with the ME model including only the photometric effects of spots is shown in Fig.~\ref{best_fit}, together with the residuals. 
%%%%%%%%%%%%%%%%%%%%%%%%%%%%%%%%%%%%%%%%%%%%%%%%%%%%%%%%%%%%%%%%%
\begin{figure}[]
\includegraphics[width=6cm,angle=90]{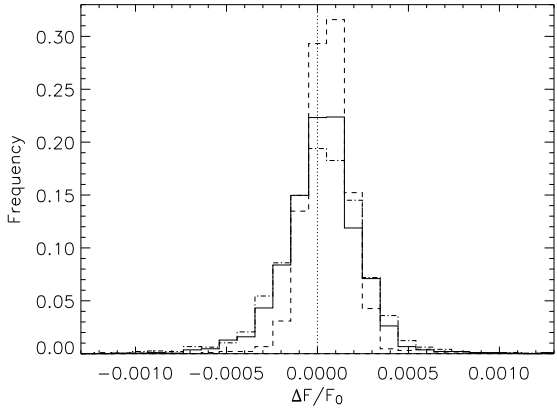} % hist_res_cp.ps 
\caption{Distribution of the residuals of the best fits of the CoRoT-Exo-2a light curve; dashed line: model with cool spots and without regularization ($\lambda=0$); solid line: ME regularized model with only cool spots; dash-dotted line: ME regularized model with cool spots and solar-like faculae. The vertical dotted line marks the zero value of the residuals. }
\label{res_distr}
\end{figure}
%%%%%%%%%%%%%%%%%%%%%%%%%%%%%%%%%%%%%%%%%%%%%%%%%%%%%%%%%%%%%%%%%

The distribution of the residuals is shown in Fig.~\ref{res_distr}, for the cases with cool spots 
without any regularization ($\lambda=0$),  with only cool spots and ME regularization, and with spots and faculae and ME regularization. The model with only cool spots and without regularization has a mean of the residuals $\mu_{\rm res}=1.27 \times 10^{-6}$ and a standard deviation $\sigma_{0}=1.44 \times 10^{-4}$ in relative flux units. The model with regularization and only cool spots has $\mu_{\rm res}=-3.29 \times 10^{-5}$ and $\sigma=2.26 \times 10^{-4}$, that including also solar-like faculae has $\mu_{\rm res}= -3.32 \times 10^{-5}$ and $\sigma=2.39 \times 10^{-4}$ in relative flux units. Note that the values of the Lagrangian multipliers have been fixed by the condition 
$|\mu_{\rm res}| = \sigma_{0} /\sqrt{N}$ where $N \sim 45$ is the number of observations in each subset of duration $\Delta t_{\rm f} =3.15611$ days. 
%%%%%%%%%%%%%%%%%%%%%%%%%%%%%%%%%%%%%%%%%%%%%%%%%%%%%%%%%%%%%%%%%
\begin{figure*}[]
\centerline{
\includegraphics[height=14cm,angle=90]{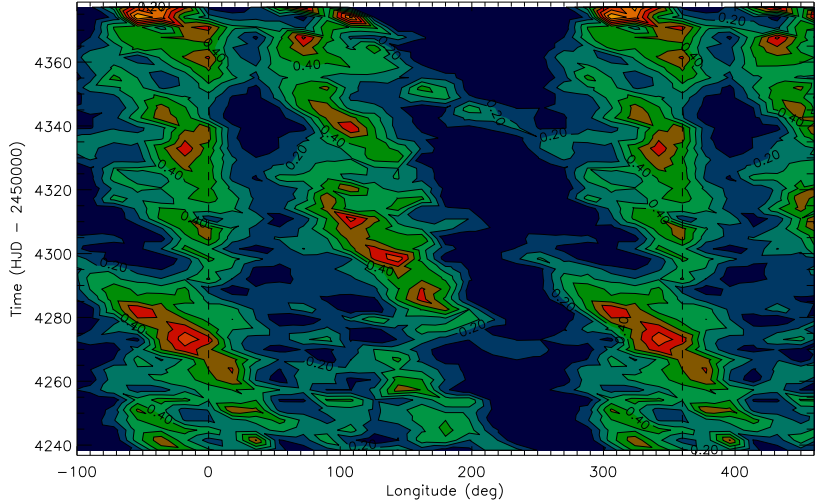} % spot_contours2.ps
} % {mag_hd179949_neu.ps}
\caption{Isocontours of the ratio $f/f_{\rm max}$, where $f$ is the spot covering factor and $f_{\rm max}= 0.0163$ its maximum value,  versus time and longitude for the ME models 
with cool spots only. The two dashed vertical lines mark longitudes $0^{\circ}$ and $360^{\circ}$ beyond which the distributions are repeated to easily follow spot migration. The contour levels are separated by $\Delta f= 0.1 f_{\rm max}$ with light yellow indicating the maximum covering factor and dark blue the minimum.}
\label{synop_spot}
\end{figure*}
%%%%%%%%%%%%%%%%%%%%%%%%%%%%%%%%%%%%%%%%%%%%%%%%%%%%%%%%%%%%%%%%%
%%%%%%%%%%%%%%%%%%%%%%%%%%%%%%%%%%%%%%%%%%%%%%%%%%%%%%%%%%%%%%%%%
\begin{figure*}[]
\centerline{
\includegraphics[height=14cm,angle=90]{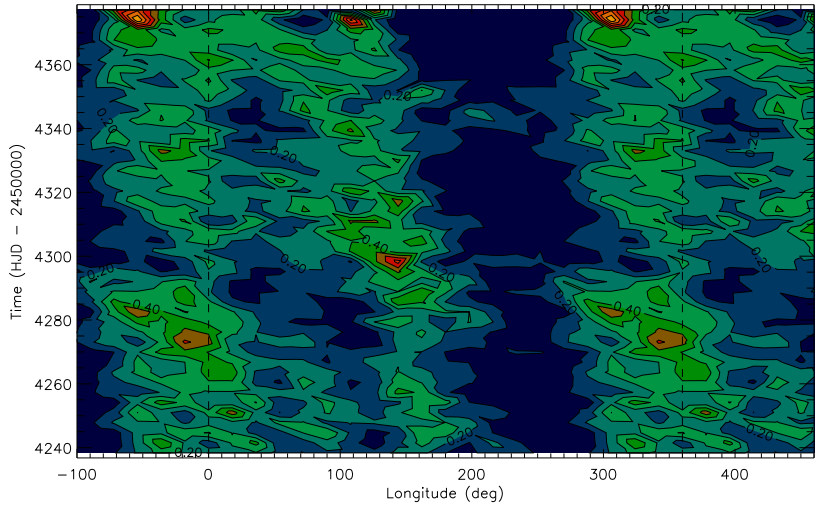} % spot_contours2f.ps 
}
\caption{The same as Fig.~\ref{synop_spot}, but for the ME models including spots and faculae with $Q=1.5$ and $f_{\rm max} = 0.0203$. }
\label{synop_sf}
\end{figure*}
%%%%%%%%%%%%%%%%%%%%%%%%%%%%%%%%%%%%%%%%%%%%%%%%%%%%%%%%%%%%%%%%%

The distribution of the spot covering factor $f$ versus longitude and time is plotted in Fig.~\ref{synop_spot}
for the ME models with spots only, and in Fig.~\ref{synop_sf} for the ME models with spots and faculae, respectively.
In both cases the longitude increases in the same direction of stellar rotation and the orbital motion of the planet.  In the case of the model with spots only, we clearly see that spots form within two active longitudes initially separated by $\sim 180^{\circ}$. The one centred around longitude $0^{\circ}$ does not appreciably migrate during the $\sim 140$ days of the observations showing a rotation period of $4.522 \pm 0.024$ days, i.e.,  the period adopted for our spot modelling,  while the other shows a retrograde migration of about $\approx 80^{\circ}$ during the same interval, which corresponds to a rotation period of 4.554 days. If such a difference corresponds to the angular velocity at different latitudes on a differentially rotating star, the relative amplitude of the differential rotation would be only 0.7 percent. { Note that the angular velocity difference between pole and equator derived from the fit in Fig.~2 of \citet{Barnesetal05} is about one order of magnitude greater, suggesting that the spots observed on CoRoT-Exo-2a are localized in a quite narrow latitude band or that our star has been observed in a state with a low degree of differential rotation \citep[cf. ][]{Donatietal03,Lanza06}.}
 
The formation and evolution of individual active regions within each of the active longitudes is clearer in Fig.~\ref{synop_spot}. It shows that an active region forms around HJD 2454245
in the active longitude centred at $0^{\circ}$ and then increases its area and migrates backward in longitude, reaching a maximum area at HJD 2454270, and eventually fading away at HJD 2454300, after $\sim 55$ days since its appearance. At HJD 2454275, a new active region forms in the other active longitude and evolves in a similar way, migrating backward in longitude and fading away at HJD 2454330, again with a characteristic  lifetime of about 55 days. Between HJD 2454330 and HJD 2454350, both active longitudes show spots with similar area that decay in $20-30$ days, leaving a minimum spotted area at HJD 2454350, when a large active region begins to develop around longitude $0^{\circ}$, reaching the maximum area at the end of the time series. The formation and evolution of individual active regions is mirrored by the beatings of the amplitude of the light modulation, as can be seen in the upper panel of Fig.~\ref{best_fit}. The minimum amplitude around HJD 2454285 can be associated with the decay of the active region around longitude $0^{\circ}$ and the rise of the other one in the second active longitude. 
The second minimum amplitude around HJD 2454350 is associated with the fading of the spots in both active longitudes immediately before the development of the large active region around longitude 
$0^{\circ}$ that dominates the light variation until the end of the observations. Therefore, the beating period of about $55$ days visible in the light curve can be associated with the typical lifetime of the largest active regions that form in the two active longitudes. { A lifetime of about two months is observed for the largest and long-lived sunspot groups and compares well with those derived from long-term photometry of active stars or the models of diffusion of  photospheric magnetic fields in stars with a low differential rotation amplitude \citep[see, e.g., ][ and references therein]{Isiketal07}.}

It is important to note that all individual spots show a remarkable retrograde migration after their formation. In other words, they have an angular velocity lower than the active longitudes by $\sim 1.3$ percent, as can be derived by the slope of the corresponding ridges in Fig.~\ref{synop_spot}.
 This behaviour is reminiscent of that of sunspot groups that show an angular velocity about 2 percent higher during the first $2-3$ days of their lifetime  than during the later stages of their evolution \citep[e.g., ][]{ZappalaZuccarello91,JavaGokhale97}. 

The ME models including the effects of faculae show a smaller relative range of variation of the spot covering factor and a systematic difference in the longitudes of the spot groups owing to the peculiar dependence of the facular contrast on the position on the stellar disc that makes their photometric effect maximum at some intermediate position between the centre and the limb, while spots have a maximum effect closer to the disc centre. Nevertheless, the general features of the models with only cool spots are reproduced, although the migration of the second active longitude initially at $180^{\circ}$ takes place at a slower pace and the  braking of individual spots is somewhat smaller. These results indicate that it is difficult to give a quantitative estimate of the surface differential rotation or evolutionary effects in the rotation of starspots in the case of CoRoT-Exo-2a owing to their significantly smaller amplitudes with respect to those of  the Sun and our limited knowledge of the impact of solar-like faculae on the observed light modulation. 

\begin{figure}[]
\centerline{
\includegraphics[width=5cm,height=8cm,angle=90]{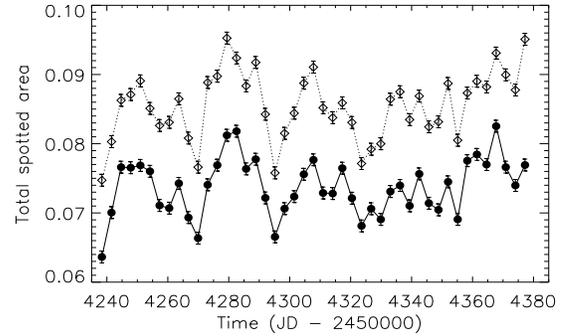} % spot_area_cp.ps} 
}
\caption{The variation of the total spotted area versus time for the ME model with spots only (filled dots; solid line) and the ME model including also the effects of faculae with $Q=1.5$ (open diamonds; dot-dashed line). The area is measured in units of the stellar photosphere.}
\label{total_spotarea}
\end{figure}

The situation is different for the variation of the total spotted area, i.e., the integral of the covering factor $f$ over the surface of the star. It is plotted versus time in 
Fig.~\ref{total_spotarea} for both the regularized models with cool spots only and with spots and faculae. {For comparison, the maximum spotted area observed in the Sun is about 0.002 of the surface}. In our model, the absolute value of the spot coverage depends on the adopted value of the unspotted magnitude, and the spot and facular contrasts, but its relative variations are largely independent of them. { In the case of the temperature deficit derived for the most active stars, (i.e., adopting $c_{\rm s}=0.21$), we find a total spotted area oscillating between 0.028 and 0.037 of the stellar surface with a variation vs. time almost identical to those in Fig.~\ref{total_spotarea}. 
Similar tests show that the shape of the modulation of the spotted area is a robust result that does not depend on the adopted (uniform) spot contrast or the inclusion of faculae in the model.} The reported error bars in 
Fig.~\ref{total_spotarea} account only for random errors in the area determination  and have a semi-amplitude of 3 standard deviations. 

\begin{figure}[]
\centerline{
\includegraphics[width=8cm,angle=0]{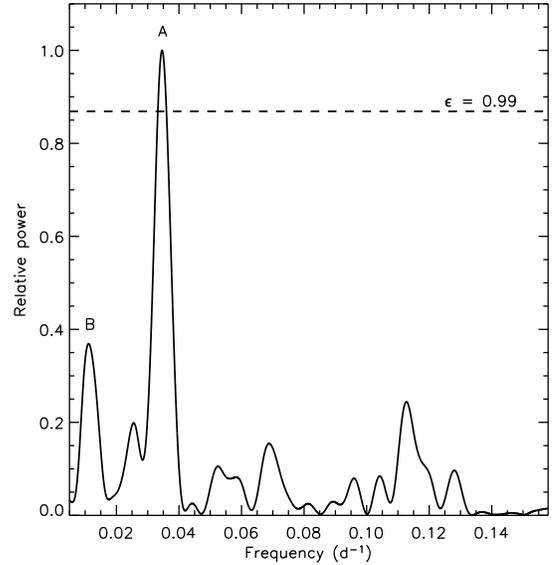} % area_per.ps 
}
\caption{Lomb-Scargle periodogram of the spotted area variation for the model including only cool spots (cf. Fig.~\ref{total_spotarea}, filled dots). The highest peak marked with A corresponds to the periodicity of 28.9 days, whereas the lower one, marked with B, corresponds to the periodicity of 86.5 days that stands out after filtering the modulation corresponding to the principal peak. The horizontal dashed line indicates the 99 percent confidence level.}
\label{area_per}
\end{figure}

 The period of the oscillations of the spotted area obtained from the model including only spots is $P_{\rm osc}= 28.9 \pm 4.8$ days with a confidence level of 99.7 percent (see Fig.~\ref{area_per}),
while for the model including also solar-like faculae we find $P_{\rm osc}= 29.5 \pm 4.8$ days with a confidence level of 99.2 percent. Note that the value of 28.9 days is close to 10 times the synodic period of the planet with respect to the rotation period of 4.522 days of the active longitudes.   
After filtering the variation at the primary frequency, we find a secondary periodicity of 86.5 days, with a conditional confidence level of only  78.3 percent \citep{HorneBaliunas86}. Its period  is three times the principal periodicity and is probably associated with the long-term variation of the mean spotted area in Fig.~\ref{total_spotarea}, although its significance is low  given the total extension of 142 days of our time series. { We investigated the  impact of temperature variations among different starspots on these results by considering two different values of the spot contrast for different longitude intervals. 
Specifically, we assumed a contrast $c_{\rm s}=0.665-\delta c_{\rm s}$ for the longitude interval from 
$54^{\circ}$ to $234^{\circ}$, roughly encompassing one of the active longitudes seen in Figs.~\ref{synop_spot} and \ref{synop_sf}, and $0.665+\delta c_{\rm s}$ for the remaining longitude range, where most of the spots of the other active longitude fall. We found that the periodicity at $\sim 29$ days is still detectable for $\delta c_{\rm} \leq 0.03$ with a confidence level larger than 95 percent. This corresponds to a relative variation of the contrast up to $\pm 4.5$ percent, i.e., about three times the standard deviation of the observed sunspot contrast at a wavelength of 672 nm 
\citep[][]{Wesolowskietal08}. Note that the relative  differences among the contrasts of individual sunspots at a given phase of the solar 11-yr cycle are greater at longer wavelengths \citep[cf., e.g.,][]{PennLivingstone06,PennMacDonald07}. However, solar monochromatic observations at $672$ nm are those falling closest to the isophotal wavelength of CoRoT observations. }  

\section{Discussion}

Our spot modelling shows that two active longitudes are present on CoRoT-Exo-2a, separated by approximately $180^{\circ}$ at the beginning of the observations. The one showing the largest concentration of spotted area rotates with a period of 4.522 days that dominates the light variation as derived from Lomb-Scargle periodogram. The other one rotates with a slightly longer period, with a relative difference in angular velocity smaller than 1 percent or even lower if we consider the models including also the effects of faculae. The possibility of using active longitudes as tracers for the surface differential rotation is supported by the recent results of \citet{BerdyuginaUsoskin03} in the case of the Sun, although the properties of solar active longitudes are still object of debate \citep[see, e.g., ][ for more details]{Usoskinetal07}. 
The angular velocity of individual spots is smaller than that of the active longitudes in CoRoT-Exo-2a, a result that suggests a dependence of their angular velocity on their evolutionary stage, as observed in the Sun. Those general results are quite robust, although a comparison between models computed without and with faculae shows that solar-like faculae have a significant impact on the determination of the spot longitudes, as already noticed by \citet{Lanzaetal07}. Therefore, our results on the amplitude of a possible surface differential rotation and the spot braking should be taken as upper limits given that they are reduced by the facular effects and the true facular area fraction $Q$ is not known in the case of CoRoT-Exo-2a. 
%%%%%%%%%%%%%%%%%%%%%%%%%%%%%%%%%%%%%%%%%%%%%%%%%%%%%%%%%%%%%%%%%
%%%%%%%%%%%%%%%%%%%%%%%%%%%%%%%%%%%%%%%%%%%%%%%%%%%%%%%%%%%%%%%%%

It is interesting to note that the braking of sunspots with their evolution can be interpreted in terms of a decrease of the depth where their flux tubes are anchored, associated with the outward decrease of the angular velocity of the Sun  in the upper layers of the solar convection zone at low latitudes \citep{SchusslerRempel05,CorbardThompson05}. In other words, our results suggest the existence of a subphotospheric shear layer with $\partial \Omega / \partial r < 0$ also in CoRoT-Exo-2a. 

\begin{figure}[]
\centerline{
\includegraphics[height=8cm,angle=90]{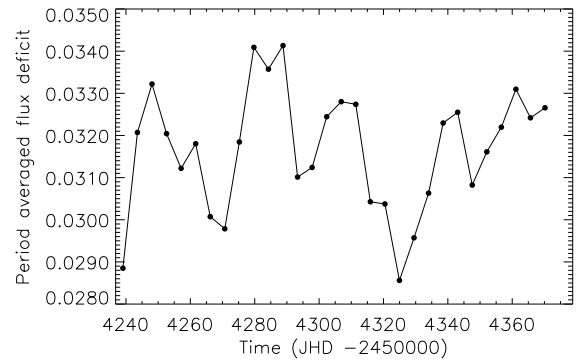} % spot_area_def.ps 
}
\caption{The relative flux deficit versus time, averaged along individual rotation periods of $4.5221$ days.}
\label{flux_deficit}
\end{figure}

On the other hand, the cyclic variation of the spotted area does not depend on the inclusion of the faculae in the model and its period can be considered as a robust result. Looking at the 
amplitude of the 
light curve modulation in Fig.~\ref{best_fit}, it is not possible to see directly the period of 28.9 days because two large spots, usually on opposite hemispheres, are modulating the stellar flux with different amplitudes. However, the periodicity is readily apparent when we plot the flux deficit, i.e., the
difference between the maximum flux and the instantaneous flux, averaged over the time along individual rotation periods of 4.52 days (see Fig.~\ref{flux_deficit}). Three cycles are clearly detectable by eye and indicate that the oscillation of the spotted area is related to an oscillation of the average light amplitude of about 0.004 mag. We have investigated whether such a periodicity may be the result of some instrumental effect by performing a Fourier analysis of the signals of the brighter stars in the asteroseismic fields, but none has been found. Moreover, we have repeated the analysis leading to the plot in Fig.~\ref{flux_deficit} for four other stars in the same exofield CCD of CoRoT-Exo-2a,  after removing long-term trends in their light curves with a low-order polynomial, and we found that none of them has the same periodicity. Their CoRoT ID, visual magnitude $V$, color index 
$B-V$ from Exodat, fraction of photometric mask contamination 
$c_{\rm m}$, angular distance $\rho $  from CoRoT-Exo-2a, and exposure time $t_{\rm e}$ are listed in Table~\ref{comparisons} together with the same parameters for our target. An independent analysis based on the method developed by \citet{Reegenetal08} confirms this result. Therefore, the cycles of the total spotted area with a period of 28.9 days are likely to be real and not related to some instrumental effect.  
%%%%%%%%%%%%%%%%%%%%%%%%%%%%%%%%%%%%%%%%%%%%%%%%%%%%%%%%%%%%%%%%%%%%%%%%%%%%%%%%
\begin{table}
\caption{Parameters of CoRoT-Exo-2a and of the comparison stars adopted to check for instrumental effects}
\begin{tabular}{lccccc}
\hline
 & & & & & \\
CoRoT ID & $V$ & $B-V$ & $c_{\rm m}$ & $\rho$ & $t_{\rm e}$ \\
 & & & & (deg) & (s) \\ 
 & & & & & \\
\hline
 & & & & & \\
CoRoT-Exo-2a & 12.568 & 0.854 & 0.0700 & 0.0 & 512, 32 \\
 101306449 & 12.788 & 0.862  &  0.0214 & 0.225 & 512  \\
101057745 & 12.690 & 0.938 & 0.0359 & 0.271 & 512,  32 \\
101297371 & 13.310 & 0.912 & 0.0654 & 0.207 & 512 \\
101275308 & 12.828 & 1.478 & 0.1712 & 0.111 & 512, 32 \\ 
& & & &  &  \\
\hline
\label{comparisons}
\end{tabular}
\end{table}
%%%%%%%%%%%%%%%%%%%%%%%%%%%%%%%%%%%%%%%%%%%%%%%%%%%%%%%%%%%%%%%%%%%%%%%%%%

The oscillation of the total spotted area with such a short period may be regarded as an analogue of the so-called solar Rieger cycles. 
\citet{Riegeretal84} reported  a periodicity of about 154 days in the occurrence of major solar flares close to the maximum of solar cycle 21 and a similar periodicity was found in the total sunspot area close to the maxima of cycles 19 and 21 \citep{Oliveretal98}. 
\citet{Lou00}  
proposed a theoretical interpretation based on Rossby-type waves propagating in the subphotospheric layers of the Sun. Their period is proportional to the rotation period of the star 
\citep[cf., e.g., Eqs. (13) and (15) in ][]{Lou00}. Therefore, the model predicts a period of about 28 days in a star rotating about five times faster than the Sun, such as CoRoT-Exo-2a. 
%%%%%%%%%%%%%%%%%%%%%%%%%%%%%%%%%%%%%%%%%%%%%%%%%%%%%%%%%%%%%%%%%
%%%%%%%%%%%%%%%%%%%%%%%%%%%%%%%%%%%%%%%%%%%%%%%%%%%%%%%%%%%%%%%%%

An alternative interpretation can be based on the coincidence of the period of 28.9 days with 10 times the synodic period of the planet with respect to the rotation of the active longitudes. This suggests  a possible star-planet interaction along the lines of the conjecture proposed by \citet{Lanza08}. Specifically, if some subphotospheric dynamo action takes place in the star, the amplified field emerges once the radial gradient of its intensity reaches some threshold value \citep{Acheson79}. The emergence of the flux might be triggered by a small localized perturbation of the dynamo  $\alpha$-effect associated with the planet \citep{Lanza08} when it passes by the  active longitudes. 
An integer number of synodic periods is necessary for the planet to trigger again field emergence in the active longitudes because the planetary perturbation can play a role  only when the field is already close to the threshold thanks to the steady amplification provided by the stellar dynamo. 

Unfortunately, the time resolution of our spot maps is only $3.15$ days, i.e., longer than the synodic period, making it impossible to resolve the effects of individual passages of the planet over each active longitude. Making the interval $\Delta t_{\rm f}$ smaller implies a loss of longitude resolution and the appearance of sizeable fluctuations of the spotted area when the modelled flux modulation covers less than half a stellar rotation. Therefore, we must limit ourselves to investigate effects on time scales longer than one synodic period. The application of eclipse mapping during successive transits may help to investigate the star-planet interaction in more detail and will be the subject of a future investigation.  
%%%%%%%%%%%%%%%%%%%%%%%%%%%%%%%%%%%%%%%%%%%%%%%%%%%%%%%%%%%%%%%%%
%%%%%%%%%%%%%%%%%%%%%%%%%%%%%%%%%%%%%%%%%%%%%%%%%%%%%%%%%%%%%%%%%

It will be interesting to observe CoRoT-Exo-2a from the ground to look for indication of star-planet magnetic interaction in the chromosphere, as recently suggested by \citet{Shkolniketal08} for some stars accompanied by hot Jupiters. 

Doppler imaging techniques can in principle be applied to CoRoT-Exo-2a because the rotational broadening of photospheric lines as measured by $v \sin i$ is $\sim 10$ km s$^{-1}$ \citep{Bouchyetal08}. Assuming a local line width of $2-2.5$ km s$^{-1}$ as a consequence of micro- and macro-turbulence, we expect to have $4-5$ resolution elements across the stellar disk which can provide a useful map with some resolution also in the latitudinal direction. However, it will be impossible to reach the uniform and extended time sampling of CoRoT photometry from the ground, even with a coordinated multisite spectroscopic campaign. Nevertheless, an independent measure of the surface differential rotation with information on the variation of the angular velocity versus latitude could be obtained with the techniques developed by \citet{Petitetal02}.

\section{Conclusions}

Our spot modelling of CoRoT-Exo-2a has provided several interesting results. Solar-like faculae have an area ratio probably significantly smaller than in the case of the Sun with $Q \sim 1.5$ giving the best fit to the light curves. Spot modelling suggests that the differential rotation on CoRoT-Exo-2a is significantly smaller than on the Sun. Adopting the relative migration of the active longitudes as a tracer for the surface shear, we find an amplitude of the relative differential rotation of $\sim 0.7$ percent when only cool spots are considered and even smaller when both spots and faculae are included in the spot modelling. 
Individual spots show a rotation rate about 1 percent smaller than that of the active longitudes where they form, a phenomenon that can be interpreted as due to the presence of a radial negative gradient of the angular velocity in the subphotospheric layers and the progressive dynamical decoupling of their magnetic flux tubes from the deep layers once spots are formed. 
A quantitative estimate of the amplitude of the spot braking is made difficult by our ignorance of the effects of the faculae on the derived spot longitudes, specifically by the lack of independent information on the $Q$ parameter. 
A robust result, that does not depend on our model assumption, is the oscillation of the total spotted area with a period of $\sim 28.9$ days during the five months of CoRoT observations. The fact that this period is close to 10 synodic periods of the planet with respect to the rotation of the active longitudes on the star, suggests some kind of magnetic interaction between the star and its hot Jupiter. Future studies of the modulation of chromospheric indicators, like the flux in the cores of Ca~\ion{II} H \& K lines, may provide further support for this interaction. 
{We note that the detection of a short-term periodicity in a  young late-type main-sequence star other than the Sun is a new result, since such short-term (or Rieger) cycles have been reported only in the RS CVn binary system UX Arietis, consisting of more evolved stars \citep{Massietal05}. } 

{The present investigation illustrates the capabilities of CoRoT  for studying the photospheric activity of late-type stars. By applying spot modelling to a large, statistically significant sample of targets, we shall be able to understand the evolution of their surface magnetic fields on time scales up to a few months and use surface brightness inhomogeneities as tracers to study differential rotation, opening a new way in the research field of solar-stellar connection.}

\begin{acknowledgements}
The authors are grateful to an anonymous Referee for a careful reading of the manuscript and several valuable comments. 
This work has been partially supported by  the Italian Space Agency (ASI) under contract  ASI/INAF I/015/07/0,
work package 3170. Active star research and exoplanetary studies at INAF-Osservatorio Astrofisico di Catania and Dipartimento di Fisica e Astronomia dell'Universit\`a degli Studi di Catania 
 is funded by MIUR ({\it Ministero dell'Istruzione, dell'Universit\`a e della Ricerca}), and by {\it Regione Siciliana}, whose financial support is gratefully
acknowledged. 
This research has made use of the ADS-CDS databases, operated at the CDS, Strasbourg, France.
\end{acknowledgements}

\end{document}